\documentclass{elsart}
\usepackage{amssymb}
\usepackage{mathrsfs}
\newtheorem{lemma}{Lemma}
\newtheorem{theorem}{Theorem}

\begin{document}

\begin{frontmatter}
\title{A lower bound on the 2-adic complexity of modified Jacobi sequence }
\author{Yuhua Sun$^{a,b,c}$, Qiang Wang$^{b}$, Tongjiang Yan$^{a,c}$}
\thanks{The work is supported by Shandong Provincial Natural Science Foundation of China (No. ZR2014FQ005), Fundamental Research Funds for the Central Universities (No. 15CX02065A, No. 15CX05060A, No. 15CX08011A, No. 15CX02056A, No. 16CX02013A, No. 16CX02009A), Fujian Provincial Key Laboratory of Network Security and Cryptology Research Fund (Fujian Normal University) (No.15002),Qingdao application research on special independent innovation plan project (No. 16-5-1-5-jch)}
\address[label1]{College of Sciences,
China University of Petroleum,
Qingdao 266555,
Shandong, China(e-mail:sunyuhua\_1@163.com), (e-mail:wang@math.carleton.ca), (e-mail:yantoji@163.com).}
\address[label2]{School of Mathematics and Statistics,
Carleton University,
Ottawa,
Ontario, K1S 5B6, Canada.}
\address[label3]{
Key Laboratory of Network Security and Cryptology,
Fujian Normal University,
Fuzhou, Fujian 350117,
China}
\begin{abstract}
Let $p,q$ be distinct primes satisfying $\mathrm{gcd}(p-1,q-1)=d$ and let $D_i$, $i=0,1,\cdots,d-1$, be Whiteman's generalized cyclotomic classes with $Z_{pq}^{\ast}=\cup_{i=0}^{d-1}D_i$. In this paper, we give the values of Gauss periods based on the generalized cyclotomic sets $D_0^{\ast}=\sum_{i=0}^{\frac{d}{2}-1}D_{2i}$ and $D_1^{\ast}=\sum_{i=0}^{\frac{d}{2}-1}D_{2i+1}$. As an application, we determine a lower bound on the 2-adic complexity of modified Jacobi sequence. Our result shows that the 2-adic complexity of modified Jacobi sequence is at least $pq-p-q-1$ with period $N=pq$. This indicates that the 2-adic complexity of modified Jacobi sequence is large enough to resist the attack of the rational approximation algorithm (RAA) for feedback with carry shift registers (FCSRs).
\end{abstract}
\end{frontmatter}
\section{Introduction}
Pseudo-random sequences with good statistical property are widely used as basic blocks for constructing stream ciphers. Any key stream generators could be implemented by both linear feedback shift registers (LFSRs) and feedback with carry shift registers (FCSRs). However, after the Berlekamp-Massey algorithm (BMA) for LFSRs \cite{Massey} and the rational approximation algorithm for FCSRs \cite{Andrew Klapper} were presented, linear complexity and 2-adic complexity of the key stream sequence have been regarded as critical security criteria and both are required to be no less than one half of the period.

Sequences from cyclotomic and generalized cyclotomic classes are large and important sequence families for constructing codebooks \cite{Hu Liqin}, \cite{Fan Cuiling}, Frequency Hopping Sequences \cite{Zeng Xiangyong}, Optical Orthogonal Codes \cite{Ding Cunsheng-4}, \cite{Ding Cunsheng-5},\cite{Cai Han} and cyclic codes \cite{Ding Cunsheng}-\cite{Ding Cunsheng-3} and most of cyclotomic sequences and generalized cyclotomic sequences have been proved to be with large linear complexity \cite{Ding Cunsheng-1}, \cite{Bai Enjian}, \cite{Yan Tongjiang-1},\cite{Li Xiaoping}. However, there are a handful research papers that focus on 2-adic complexities of these sequences. In fact, although the concept of 2-adic has been presented for more than two decades, there are only a few kinds of sequences whose 2-adic complexities have been completely determined. For example, in 1997, Klapper has pointed out that an $m$-sequence with prime period has maximal 2-adic complexity \cite{Andrew Klapper}. In 2010, Tian and Qi showed that the 2-adic complexity of all the binary $m$-sequences is maximal \cite{Tian Tian}. Afterwards, Xiong et al. \cite{Xiong Hai} presented a new method using circulant matrices to compute the 2-adic complexities of binary sequences. They showed that all the known sequences with ideal 2-level autocorrelation have maximum 2-adic complexity. Moreover, they also proved that the 2-adic complexities of Legendre sequences and Ding-Helleseth-Lam sequences
 with optimal autocorrelation are also maximal. Then, using the same method as that in \cite{Xiong Hai}, Xiong et al. \cite{Xiong Hai-2} pointed out that two other classes of sequences based on interleaved structure have also maximal 2-adic complexity. One of these two classes of sequences was constructed by Tang and Ding \cite{Tang-Ding}, which has optimal autocorrelation, the other was constructed by Zhou et al \cite{Zhou}, which is optimal with respect to the Tang-Fan-Matsufuji bound \cite{Tang-Fan bound}.

Modified Jacobi sequence is one of sequence families constructed by Whiteman generalized cyclotomic classes. Green and Choi have proved that these sequences have large linear complexity and low autocorrelation in many cases. But, as far as the authors known, among Jacobi sequence family, there is no other result about the 2-adic complexities of these sequences other than twin-prime sequences, which has ideal autocorrelation and has been proved to be with maximal 2-adic complexity by Xiong et al \cite{Xiong Hai}, and another class of sequences constructed from Whiteman generalized cyclotomic classes of order 2 has been proved to be also with maximal 2-adic complexity by Zeng et al. \cite{Zeng Xiangyong}.

In this paper, we study the 2-adic complexity of modified Jacobi sequences. And we give a general lower bound on the 2-adic complexity, i.e., we will prove that the 2-adic complexity of all these sequences is lower bounded by $pq-p-q-1$ with period $N=pq$. As a special case, we can also confirm that the twin-prime sequence has maximal 2-adic complexity.

The rest of this paper is organized as follows. Some necessary definitions, notations,  and previous results are introduced in Section 2. Gauss periods based on two Whiteman generalized cyclic sets are given in section 3. And the lower bound on the 2-adic of modified Jacobi sequence is given in Section 4. Finally we  summarize our results and give some remarks in Section 5.
\section{Preliminaries}\label{section 2}
Let $N$ be a positive integer, $\{s_i\}_{i=0}^{N-1}$ a binary sequence of period $N$, and $S(x)=\sum\limits_{i=0}^{N-1}s_{i}x^i\in \mathbb{Z} [x]$. If we write
\begin{equation}
\frac{S(2)}{2^N-1}=\frac{\sum\limits_{i=0}^{N-1}s_{i}2^i}{2^N-1}=\frac{m}{n},\ 0\leq m\leq n,\ \mathrm{gcd}(m,n)=1,\label{2-adic complexity}
\end{equation}
then the 2-adic complexity $\Phi_{2}(s)$ of the sequence $\{s_i\}_{i=0}^{N-1}$ is defined as the integer $\lfloor\mathrm{log}_2n\rfloor$, i.e.,
\begin{equation}
\Phi_{2}(s)=\left\lfloor\mathrm{log}_2\frac{2^N-1}{\mathrm{gcd}(2^N-1,S(2))}\right\rfloor,\label{2-adic calculation}
\end{equation}
where $\lfloor x\rfloor$ is the greatest integer that is less than or equal to $x$.

Let $p=df+1$ and $q=df^{\prime}+1$ ($p<q$) be two odd primes with $\gcd(p-1, q-1)=d$.
Define $N=pq$, $\mathcal{L}=(p-1)(q-1)/d$. The Chinese Remainder Theorem
guarantees that there exists a common primitive root $g$ of both $p$
and $q$. Then the order of $g$ modulo $N$ is $\mathcal{L}$. Let $x$ be an
integer satisfying $x\equiv g\pmod{p},\  x\equiv1\pmod{q}.$
The existence and uniqueness of $x\pmod{N}$ are also guaranteed by
the Chinese Remainder Theorem. Whiteman \cite{Whiteman-1} presented the definition of the following generalized cyclotomic classes
\begin{equation}
D_i=\{g^{t}x^i: t=0, 1, \cdots, \mathcal{L}-1\},\ i=0, 1, \cdots, d-1\label{cyclotomic class}
\end{equation}
of order $d$. And he also has proved
$$Z_N^{\ast}=\cup_{i=0}^{d-1}D_i,\ D_i\cap D_j=\emptyset\,\ \mbox{for}\ i\neq j,$$
where $\emptyset$ denotes the
empty set. The corresponding generalized cyclotomic numbers of order $d$ are defined by
$$(i,j)=|(D_{i}+1)\cap D_{j}|,\ \mathrm{for\ all}\ i,j=0,1,\cdots, d-1.$$
And the following properties of the generalized cyclotomic numbers have also been given by Whiteman:
\begin{eqnarray}
(i,j)&=&(d-i,j-i),\label{prop-1}\\
(i,j)&=&\left\{
\begin{array}{ll}
(j+\frac{d}{2},i+\frac{d}{2}),\ \mathrm{ if}\ ff^{\prime}\ \mathrm{is\ even},\\
(j,i),\ \ \ \ \ \ \ \ \ \ \ \ \ \mathrm{if}\ ff^{\prime}\ \mathrm{is\ odd},
\end{array}
\right.\label{prop-2}\\
\sum_{j=0}^{d-1}(i,j)&=&\frac{(p-2)(q-2)-1}{d}+\delta_i,\label{prop-3}
\end{eqnarray}
where
$$
\delta_i=\left\{
\begin{array}{ll}
1,\ \mathrm{if}\ ff^{\prime}\ \mathrm{is\ even\ and}\ i=\frac{d}{2},\mathrm{\ or\ if\ }ff^{\prime}\ \mathrm{is\ odd\ and}\ i=0,\\
0,\ \mathrm{otherwise.}
\end{array}
\right.
$$
If we denote
$$
\begin{array}{lll}
  P=\{p, 2p, \cdots, (q-1)p\},\
  Q=\{q, 2q, \cdots, (p-1)q\},\
  R=\{0\},\\
C_{0}=\cup_{i=0}^{\frac{d}{2}-1}D_{2i}\cup Q\cup R,\ C_{1}=\cup_{i=0}^{\frac{d}{2}-1}D_{2i+1}\cup P,\\
D_0^{\ast}=\cup_{i=0}^{\frac{d}{2}-1}D_{2i},\ D_1^{\ast}=\cup_{i=0}^{\frac{d}{2}-1}D_{2i+1},
\end{array}
$$
then we have $\mathbb{Z}_N=C_0\cup C_{1}$, $\mathbb{Z}_N^{\ast}=D_0^{\ast}\cup D_1^{\ast}$ and it is easy to see that $D_0^{\ast}$ is a subgroup of $\mathbb{Z}_N^{\ast}$. The modified Jacobi sequence $\{s_i\}_{i=0}^{N-1}$ is defined by
\begin{equation}\label{definition}
s_i=\left\{
\begin{array}{ll}
0, \ \mathrm{if}\ i\pmod{N}\in C_0,\\
1, \ \mathrm{if}\ i\pmod{N}\in C_1.
\end{array}
\right.
\end{equation}
Moreover, it is not difficult to verify that modified Jacobi sequence is also equivalent to the following definition
$$
s_i=\left\{
\begin{array}{llll}
1,\ \ \ \ \ \ \ \ \ \ \mathrm{if}\ i\pmod{N}\in P,\\
\frac{1-\left(\frac{i}{p}\right)\left(\frac{i}{q}\right)}{2},\ \mathrm{if}\ i\in Z_{N}^{\ast},\\
0,\ \ \ \ \ \ \ \ \ \ \mathrm{otherwise},
\end{array}
\right.
$$
where $(\cdot)$ is the Legendre symbol. But, in this paper, we will discuss its 2-adic complexity using properties of generalized cyclotomic classes.

From the above argument, we have known that $D_0^{\ast}$ is a subgroup of $\mathbb{Z}_{N}^{\ast}$, then $D_0^{\ast}$ and $D_1^{\ast}$ are also generalized cyclotomic classes. In this paper, we will determine the Gauss periods based on $D_0^{\ast}$ and $D_1^{\ast}$. Then, using these gauss periods, a lower bound on the 2-adic of modified Jacobi sequence will be given. To this end, we will first list the following properties of the above sets (The proofs of these properties can also be found in many literatures, for example, see \cite{Yan-thesis}).
\begin{lemma}\label{basic cyclotomic-1}
For any $a\in \mathbb{Z}_N$ and $B\subseteq \mathbb{Z}_N$, we denote $aB=\{ab|b\in B\}$. Then we have the following properties.
\begin{itemize}
\item[(1)]For each fixed $a\in D_i$, we have $aD_j=D_{(i+j)\pmod{d}}$, $aP=P$ and $aQ=Q$, where $i,j=0,1,\cdots,d-1$.
\item[(2)] For each fixed $a\in P$, if $b$ runs through each element of $D_i$, $i=0,1,\cdots,d-1$, then $ab$ runs exactly each element of $P$ $\frac{p-1}{2}$ times. Symmetrically, for each fixed $a\in Q$, if $b$ runs through each element of $D_i$, $i=0,1,\cdots,d-1$, then $ab$ runs exactly each element of $Q$ $\frac{q-1}{2}$ times.
\item[(3)]For each fixed $a\in P$, we have $aP=P$, $aQ=R$. Symmetrically, for each fixed $a\in Q$, we have $aQ=Q$, $aP=R$.
\item[(4)]For each fixed $a\in D_i^{\ast}$, we have $aD_j^{\ast}=D_{(i+j)\pmod{2}}^{\ast}$, $aP=P$ and $aQ=Q$, where $i,j=0,1$.
\end{itemize}
\end{lemma}

Let $\omega_N=e^{2\pi\sqrt{-1}/N}$ be a $N$th complex primitive root of unity. Then the additive character $\chi$ of $\mathbb{Z}_N$ is given by
\begin{equation}
\chi(x)=\omega_N^{x},\ x\in \mathbb{Z}_N\label{character}
\end{equation}
and Gaussian periods of order $d$ are defined by
$$
\eta_{i}=\sum_{x\in D_{i}}\chi(x),\ i=0,1,\cdots,d-1.
$$
It is well-known that
\begin{equation}
\sum_{x\in P}\chi(x)=-1,\label{basic gaussian sum-2}
\end{equation}
\begin{equation}
\sum_{x\in Q}\chi(x)=-1,\label{basic gaussian sum-3}
\end{equation}
and
\begin{equation}
\sum_{i=0}^{d-1}\eta_i=1.\label{basic gaussian sum}
\end{equation}
Moreover, the following results, which have also been proved by Whiteman \cite{Whiteman-1}, will be useful.
\begin{lemma}\label{-1 belong to}
The element $-1\in \mathbb{Z}_{N}^{\ast}$ satisfies
$$
-1\equiv\left\{
\begin{array}{ll}
g^{\delta}x^{\frac{d}{2}}\pmod{N},\ \mathrm{if}\ ff^{\prime}\ \mathrm{is\ even},\\
g^{\frac{\mathcal{L}}{2}}\pmod{N},\ \ \ \ \mathrm{if}\ ff^{\prime}\ \mathrm{is\ odd},
\end{array}
\right.
$$
where $\delta$ is some fixed integer such that $0\leq \delta\leq \mathcal{L}-1$
\end{lemma}
\begin{lemma}\label{cyclotomic numbers-1}
For each $u\in P\cup Q$,
$$
|D_i\cap(D_j+u)|=\left\{
\begin{array}{lll}
\frac{(p-1)(q-1)}{d^{2}},\ \ \ \ \mathrm{if}\ i\neq j,\\
\frac{(p-1)(q-1-d)}{d^{2}},\ \mathrm{if}\ i=j\ and\ u\in P,\\
\frac{(p-1-d)(q-1)}{d^{2}},\ \mathrm{if}\ i=j\ and\ u\in Q.
\end{array}
\right.
$$
\end{lemma}
\section{Gaussian periods of Whiteman generalized cyclotomic classes}
Let $\Omega_0=\sum_{x\in D_0^{\ast}}\chi(x)=\sum_{i=0}^{\frac{d}{2}-1}\eta_{2i}$ and $\Omega_1=\sum_{x\in D_1^{\ast}}\chi(x)=\sum_{i=0}^{\frac{d}{2}-1}\eta_{2i+1}$. In this section, we will determine the values of $\Omega_0$ and $\Omega_1$.
\begin{theorem}\label{gauss period-thm}
Let $p=df+1,q=df^{\prime}$ be distinct primes satisfying $\mathrm{gcd}(p-1,q-1)=d$ and let $D_i^{\ast}$ be the generalized cyclotomic set be defined in section 2 and $\Omega_0$ the gauss period based on $D_i^{\ast}$, where $i=0,1$. Then the Gauss periods are given by
\begin{eqnarray}
\Omega_{0}&=&\sum_{x\in D_0^{\ast}}\chi(x)=\left\{
\begin{array}{lll}
\frac{1\pm\sqrt{pq}}{2},\ \ \mathrm{if}\ ff^{\prime}\ \mathrm{is\ odd},\ \mathrm{or\ if}\\
\ \ \ \ \ \ \ \ \ ff^{\prime}\ \mathrm{is\ even\ and}\ d\equiv0\pmod{4},\\
\frac{1\pm\sqrt{-pq}}{2}, \mathrm{if}\ ff^{\prime}\ \mathrm{is\ even\ and}\ d\equiv2\pmod{4},\\
\end{array}
\right.\label{gauss period-1}\\
\Omega_{1}&=&1-\Omega_{0}.\label{gauss period-2}
\end{eqnarray}
\end{theorem}
\noindent{\bf Proof.} Above all, from the definition of generalized cyclotomic class, for any $\tau\in D_k$, it can be easy verify that $\tau^{-1}\in D_{(d-k)\ (\mathrm{mod}\ d)}$. Then, by Lemma \ref{basic cyclotomic-1}, for any $0\leq i,j,k\leq d-1$, we have
\begin{eqnarray}
\mid(D_i+\tau)\cap D_j\mid&=&\mid(\tau^{-1}D_i+1)\cap \tau^{-1}D_j\mid\nonumber\\
&=&\left((i+d-k)\ (\mathrm{mod}\ d),(j+d-k)\ (\mathrm{mod}\ d)\right).\nonumber
\end{eqnarray}
If $ff^{\prime}$ is odd, then, by Lemma \ref{-1 belong to}, we have $-1\in D_0$. Therefore, using Lemma \ref{cyclotomic numbers-1}, we get
\begin{eqnarray}
\left(\Omega_0\right)^2&=&\left(\sum_{i=0}^{\frac{d}{2}-1}\eta_{2i}\right)^2=\sum_{i=0}^{\frac{d}{2}-1}\eta_{2i}^2+\sum_{i=0}^{\frac{d}{2}-1}\sum_{j\neq i, j=0}^{\frac{d}{2}-1}\eta_{2i}\eta_{2j}\nonumber\\
&=&\sum_{i=0}^{\frac{d}{2}-1}\left(\sum_{x\in D_{2i}}\sum_{y\in D_{2i}}\omega_{N}^{y-x}\right)+\sum_{i=0}^{\frac{d}{2}-1}\sum_{j\neq i, j=0}^{\frac{d}{2}-1}\left(\sum_{x\in D_{2i}}\sum_{y\in D_{2j}}\omega_{N}^{y-x}\right)\nonumber\\
&=&\sum_{i=0}^{\frac{d}{2}-1}\left(\sum_{k=0}^{d-1}(2i-k,2i-k)\eta_k+\frac{(p-1)(q-1)}{d}-\frac{(p-1)(q-1-d)}{d^2}\right.\nonumber\\
&-&\left.\frac{(p-1-d)(q-1)}{d^2}\right)\nonumber\\
&+&\sum_{i=0}^{\frac{d}{2}-1}\sum_{j\neq i,j=0}^{\frac{d}{2}-1}\left(\sum_{k=0}^{d-1}(2i-k,2j-k)\eta_k-\frac{(p-1)(q-1)}{d^{2}}-\frac{(p-1)(q-1)}{d^{2}}\right)\nonumber\\
&=&\sum_{k=0}^{d-1}\left(\sum_{i=0}^{\frac{d}{2}-1}\sum_{j=0}^{\frac{d}{2}-1}(2i-k,2j-k)\right)\eta_k+\frac{p-1}{2}+\frac{q-1}{2}\nonumber\\
&=&\sum_{k=0}^{d-1}\left(\sum_{i=0}^{\frac{d}{2}-1}\sum_{j=0}^{\frac{d}{2}-1}(d-(2i-k),2(j-i))\right)\eta_k+\frac{p-1}{2}+\frac{q-1}{2}\label{ref-of property}\\
&=&\sum_{k=0}^{d-1}\left(\sum_{j^{\prime}=0}^{\frac{d}{2}-1}\sum_{j=0}^{\frac{d}{2}-1}(d-(2(j-j^{\prime})-k),2j^{\prime})\right)\eta_k+\frac{p-1}{2}+\frac{q-1}{2},\nonumber
\end{eqnarray}
where Eq. (\ref{ref-of property}) comes from Eq.(\ref{prop-1}).
Similarly, we can get
\begin{eqnarray}
\left(\Omega_1\right)^2&=&\left(\sum_{i=0}^{\frac{d}{2}-1}\eta_{2i+1}\right)^2=\sum_{i=0}^{\frac{d}{2}-1}\eta_{2i+1}^2+\sum_{i=0}^{\frac{d}{2}-1}\sum_{j\neq i,j=0}^{\frac{d}{2}-1}\eta_{2i+1}\eta_{2j+1}\nonumber\\
&=&\sum_{k=0}^{d-1}\left(\sum_{j^{\prime}=0}^{\frac{d}{2}-1}\sum_{j=0}^{\frac{d}{2}-1}\left(d-(2(j-j^{\prime})+1-k),2j^{\prime}\right)\right)\eta_k+\frac{p-1}{2}+\frac{q-1}{2}.\nonumber
\end{eqnarray}
Then,
\begin{eqnarray}
\left(\Omega_0\right)^2+\left(\Omega_1\right)^2&=&\sum_{k=0}^{d-1}\left(\sum_{j^{\prime}=0}^{\frac{d}{2}-1}\left(\sum_{j=0}^{\frac{d}{2}-1}
\left((d-(2(j-j^{\prime})-k),2j^{\prime})\right.\right.\right.\nonumber\\
&+&\left.\left.\left.(d-(2(j-j^{\prime})+1-k),2j^{\prime})\right)\right)\right)\eta_k\nonumber\\
&+&(p-1)+(q-1)\nonumber\\
&=&\sum_{k=0}^{d-1}\left(\sum_{j^{\prime}=0}^{\frac{d}{2}-1}\left(\sum_{j=0}^{d-1}
(j,2j^{\prime})\right)\right)\eta_k+(p-1)+(q-1)\nonumber\\
&=&\sum_{j^{\prime}=0}^{\frac{d}{2}-1}\left(\sum_{j=0}^{d-1}
(j,2j^{\prime})\right)+(p-1)+(q-1)\label{ref-gauss basic}\\
&=&\sum_{j=0}^{d-1}
(j,0)+\sum_{j^{\prime}=1}^{\frac{d}{2}-1}\left(\sum_{j=0}^{d-1}
(j,2j^{\prime})\right)+(p-1)+(q-1)\nonumber\\
&=&\sum_{j=0}^{d-1}
(0,j)+\sum_{j^{\prime}=1}^{\frac{d}{2}-1}\left(\sum_{j=0}^{d-1}
(2j^{\prime},j)\right)+(p-1)+(q-1)\label{ref-prop-1}\\
&=&1+\frac{d}{2}\times\frac{(p-2)(q-2)-1}{d}+(p-1)+(q-1)\label{ref-property}\\
&=&1+\frac{pq-1}{2},\nonumber
\end{eqnarray}
Where Eq. (\ref{ref-gauss basic}) is by Eq. (\ref{basic gaussian sum}), Eq. (\ref{ref-prop-1}) is by Eq. (\ref{prop-2}), and Eq. (\ref{ref-property}) hold because of Eq. (\ref{prop-3}).
Moreover, from Eq.(\ref{basic gaussian sum}), we know that $\Omega_{0}+\Omega_1=1$ and
\begin{eqnarray}
\left(\Omega_0\right)^2+\left(\Omega_1\right)^2&=&\left(\Omega_{0}+\Omega_1\right)^2-2\Omega_0\Omega_1\nonumber\\
&=&1-2\Omega_0\Omega_1\nonumber
\end{eqnarray}
Then we obtain $\Omega_0\Omega_1=\frac{1-pq}{4}$, which implies
$\Omega_0=\frac{1\pm\sqrt{pq}}{2}$ and $\Omega_1=\frac{1\mp\sqrt{pq}}{2}$.

Suppose that $ff^{\prime}$ is even then, by Lemma \ref{-1 belong to}, we have $-1\in D_{\frac{d}{2}}$. Let $d\equiv2\pmod{4}$, i.e., $\frac{d}{2}$ is odd. Then, for any $0\leq i,j\leq\frac{d}{2}-1$ with $i\neq j$, we know that $2i+\frac{d}{2}\not\equiv2j\pmod{d}$ and $2i+1+\frac{d}{2}\not\equiv2j+1\pmod{d}$. Therefore,
\begin{eqnarray}
\left(\Omega_0\right)^2&=&\sum_{i=0}^{\frac{d}{2}-1}\sum_{j=0}^{\frac{d}{2}-1}\left(\sum_{x\in D_{2i+\frac{d}{2}}}\sum_{y\in D_{2j}}\omega_{N}^{y-x}\right)\nonumber\\
&=&\sum_{i=0}^{\frac{d}{2}-1}\sum_{j=0}^{\frac{d}{2}-1}\left(\sum_{k=0}^{d-1}(2i+\frac{d}{2}-k,2j-k)\eta_k-2\times\frac{(p-1)(q-1)}{d^2}\right)\nonumber\\
&=&\sum_{k=0}^{d-1}\left(\sum_{i=0}^{\frac{d}{2}-1}\sum_{j=0}^{\frac{d}{2}-1}(\frac{d}{2}-(2i-k),2(j-i)+\frac{d}{2})\right)\eta_k-\frac{(p-1)(q-1)}{2}\label{ref-property-3-0}\\
&=&\sum_{k=0}^{d-1}\left(\sum_{i=0}^{\frac{d}{2}-1}\sum_{j=0}^{\frac{d}{2}-1}(2(j-i),k-2i)\right)\eta_k-\frac{(p-1)(q-1)}{2}\label{ref-property-3}\\
&=&\sum_{k=0}^{d-1}\left(\sum_{j^{\prime}=0}^{\frac{d}{2}-1}\sum_{j=0}^{\frac{d}{2}-1}(2j^{\prime},k-2(j-j^{\prime}))\right)\eta_k-\frac{(p-1)(q-1)}{2},\nonumber
\end{eqnarray}
where Eq. (\ref{ref-property-3-0}) comes again from Eq. (\ref{prop-1}) and Eq. (\ref{ref-property-3}) comes from Eq. (\ref{prop-2}).
Similarly, we can get
\begin{eqnarray}
\left(\Omega_1\right)^2&=&\sum_{k=0}^{d-1}\left(\sum_{j^{\prime}=0}^{\frac{d}{2}-1}\sum_{j=0}^{\frac{d}{2}-1}(2j^{\prime},k-2(j-j^{\prime})-1)\right)\eta_k-\frac{(p-1)(q-1)}{2}.\nonumber
\end{eqnarray}
Then we get
\begin{eqnarray}
\left(\Omega_0\right)^2+\left(\Omega_1\right)^2&=&\sum_{k=0}^{d-1}\left(\sum_{j^{\prime}=0}^{\frac{d}{2}-1}\left(\sum_{j=0}^{\frac{d}{2}-1}
\left((2j^{\prime},k-2(j-j^{\prime}))+(2j^{\prime},k-2(j-j^{\prime})-1)\right)\right)\right)\eta_k\nonumber\\
&-&(p-1)(q-1)\nonumber\\
&=&\sum_{k=0}^{d-1}\left(\sum_{j^{\prime}=0}^{\frac{d}{2}-1}\left(\sum_{j=0}^{d-1}
(2j^{\prime},j)\right)\right)\eta_k-(p-1)(q-1)\nonumber\\
&=&\sum_{j^{\prime}=0}^{\frac{d}{2}-1}\left(\sum_{j=0}^{d-1}
(2j^{\prime},j)\right)-(p-1)(q-1)\label{ref-prop-3}\\
&=&\frac{d}{2}\times\frac{(p-2)(q-2)-1}{d}-(p-1)(q-1)\label{ref-prop-2}\\
&=&1-\frac{pq+1}{2},\nonumber
\end{eqnarray}
where Eq. (\ref{ref-prop-3}) is by Eq. (\ref{basic gaussian sum}) and Eq. (\ref{ref-prop-2}) is by Eq. (\ref{prop-3}).
Similar argument to the above, we can get $\Omega_0\Omega_1=\frac{1+pq}{4}$, which implies
$\Omega_0=\frac{1\pm\sqrt{-pq}}{2}$ and $\Omega_1=\frac{1\mp\sqrt{-pq}}{2}$.
For the case of even $ff^{\prime}$ and $d\equiv0\pmod{4}$, the result can be similarly obtained.
\ \ \ \ \ \ \ \ \ \ \ \ \ \ \ \ \ \ \ \ \ \ \ \ \ \ \ \ \ \ \ \ \ \ \ \ \ \ \ \ \ \ \ \ \ \ \ \ \ \ \ \ \ \ \ \ \ \ \ \ \ \ \ \ \ \ \ \ \ \ \ \ \ \ \ \ \ \ \ \ \ \ \ \ \ \ $\Box$

\section{A lower bound on the 2-adic complexity of Jacobi sequence} \label{section 4}

Let $\{s_i\}_{i=0}^{N-1}$ be a binary sequence with period $N$ and let $A=(a_{i,j})_{N\times N}$ be the matrix defined by $a_{i,j}=s_{i-j\pmod{N}}$. In this section, using the periods which have been determined in Section 3, we will give a lower bound on the 2-adic complexity of modified Jacobi sequence. In order to derive the lower bound, the following two results will be useful, which can be found in \cite{Xiong Hai} and \cite{Davis} respectively.

\begin{lemma}\label{main method-1}\cite{Xiong Hai}
Viewing $A$ as a matrix over the rational fields $\mathbb{Q}$, if $\mathrm{det}(A)\neq0$, then
\begin{equation}
\mathrm{gcd}\left(S(2),2^N-1\right)|\mathrm{gcd}\left(\mathrm{det}(A),2^N-1\right).\label{method-1}
\end{equation}
\end{lemma}

\begin{lemma}\label{main method-2}\cite{Davis}
 $\mathrm{det}(A)=\prod_{a=0}^{N-1}S(\omega_N^a)$, where $\omega_N$ is defined as in Eq. (\ref{character}).
\end{lemma}
\begin{lemma}\label{main method-3}
Let $\{s_i\}_{i=0}^{N-1}$ be the modified Jacobi sequence with period $N=pq$. Then we have
\begin{equation}
S(\omega_{N}^{a})=\left\{
\begin{array}{lllll}
\frac{(p+1)(q-1)}{2},\ \mathrm{if}\ a\in R,\\
-\Omega_0,\ \ \ \ \ \ \ \mathrm{if}\ a\in D_0^{\ast},\\
-\Omega_1,\ \ \ \ \ \ \ \mathrm{if}\ a\in D_1^{\ast},\\
-\frac{p+1}{2},\ \ \ \ \ \ \mathrm{if}\ a\in P,\\
\ \frac{q-1}{2},\ \ \ \ \ \ \ \mathrm{if}\ a\in Q.
\end{array}
\right.
\end{equation}
\end{lemma}
\noindent{\bf Proof.}
Recall that
\begin{eqnarray}
S(\omega_N^a)&=&\Sigma_{k\in C_1}(\omega_N^a)^k=\Sigma_{k\in D_1^{\ast}}(\omega_N^a)^k+\Sigma_{k\in P}(\omega_N^a)^k\nonumber\\
&=&\Sigma_{k\in aD_1^{\ast}}\omega_N^k+\Sigma_{k\in aP}\omega_N^k.\nonumber
\end{eqnarray}
Firstly, if $a=0$, then it is easy to see $S(\omega_N^a)=\frac{(p-1)(q-1)}{2}+(q-1)=\frac{(p+1)(q-1)}{2}$.
Secondly, if $a\in D_0^{\ast}$, by Lemma \ref{basic cyclotomic-1} and Eq. (\ref{basic gaussian sum-2}), then we have
 \begin{eqnarray}
S(\omega_N^a)&=&\Sigma_{k\in aD_1^{\ast}}\omega_N^k+\Sigma_{k\in aP}\omega_N^k=\Sigma_{k\in D_1^{\ast}}\omega_N^k+\Sigma_{k\in P}\omega_N^k\nonumber\\
&=&\Omega_1-1=-\Omega_0\ \ \ (\mathrm{for}\ \Omega_0+\Omega_1=1\ \mathrm{by\ Eq}. (\ref{basic gaussian sum})).\nonumber
\end{eqnarray}
Similarly, if $a\in D_1^{\ast}$, we have
 \begin{eqnarray}
S(\omega_N^a)&=&\Sigma_{k\in aD_1^{\ast}}\omega_N^k+\Sigma_{k\in aP}\omega_N^k=\Sigma_{k\in D_0^{\ast}}\omega_N^k+\Sigma_{k\in P}\omega_N^k\nonumber\\
&=&\Omega_0-1=-\Omega_1.\nonumber
\end{eqnarray}
If $a\in P$, again by Lemma \ref{basic cyclotomic-1}, we have
 \begin{eqnarray}
S(\omega_N^a)&=&\Sigma_{k\in aC_1}\omega_N^k+\Sigma_{k\in aP}\omega_N^k=\frac{p-1}{2}\Sigma_{k\in P}\omega_N^k+\Sigma_{k\in P}\omega_N^k\nonumber\\
&=&-\frac{p-1}{2}-1=-\frac{p+1}{2}.\nonumber
\end{eqnarray}
Similarly, if $a\in Q$ we have
 \begin{eqnarray}
S(\omega_N^a)&=&\Sigma_{k\in aC_1}\omega_N^k+\Sigma_{k\in aP}\omega_N^k=\frac{q-1}{2}\Sigma_{k\in Q}\omega_N^k+(q-1)\nonumber\\
&=&-\frac{q-1}{2}+(q-1)=\frac{q-1}{2}.\nonumber
\end{eqnarray}
The result follows. \ \ \ \ \ \ \ \ \ \ \ \ \ \ \ \ \ \ \ \ \ \ \ \ \ \ \ \ \ \ \ \ \ \ \ \ \ \ \ \ \ \ \ \ \ \ \ \ \ \ \ \ \ \ \ \ \ \ \ \ \ \ \ \ \ \ \ \ \ \ \ \ \ \ \ \ \ \ \ \ \ \  \ \ \ \ $\Box$

\begin{lemma}\label{num-theory}
Let $p$ and $q$ be two distinct odd primes and $N=pq$. Then we have $\mathrm{gcd}(2^p-1,\frac{2^N-1}{2^p-1})=\mathrm{gcd}(2^p-1,q)$ and $\mathrm{gcd}(2^q-1,\frac{2^N-1}{2^q-1})=\mathrm{gcd}(2^q-1,p)$. Particularly, if $q>p$, we have $\mathrm{gcd}(2^q-1,p)=1$, i.e., $\mathrm{gcd}(2^q-1,\frac{2^N-1}{2^q-1})=1$.
\end{lemma}
\noindent{\bf Proof.}
Note that
\begin{eqnarray}
2^N-1&=&2^{pq}-1=(2^p-1)(2^{p(q-1)}+2^{p(q-2)}+\cdots+2^p+1)\nonumber\\
&=&2^{pq}-1=(2^q-1)(2^{q(p-1)}+2^{q(p-2)}+\cdots+2^q+1).\nonumber
\end{eqnarray}
Then we get $\frac{2^{pq}-1}{2^p-1}\equiv q\pmod{2^p-1}$ and $\frac{2^{pq}-1}{2^q-1}\equiv p\pmod{2^q-1}$, which imply
$\mathrm{gcd}(2^p-1,\frac{2^N-1}{2^p-1})=\mathrm{gcd}(2^p-1,q)$ and $\mathrm{gcd}(2^q-1,\frac{2^N-1}{2^q-1})=\mathrm{gcd}(2^q-1,p)$.
Particularly, for $q>p$, if $\mathrm{gcd}(2^q-1,p)>1$, i.e., $p|2^q-1$, then the multiplicative order of 2 modular $p$, denoted as $\mathrm{Ord}_p(2)$, is a divisor of $q$. But $q$ is a prime, then $\mathrm{Ord}_p(2)=q$. By Fermat Theorem, we know that $p|2^{p-1}-1$. Therefore, we have $q\leq p-1$, which contradicts to the fact $p<q$. The desired result follows.
\ \ \ \ \ \ \ \ \ \ \ \ \ \ \ \ \ \ \ \ \ \ \ \ \ \ \ \ \ \ \ \ \ \ \ \ \ \ \ \ \ \ \ \ \ \ \ \ \ \ \ \ \ \ \ \ \ \ \ \ \ \ \ \ \ \ \ \ \ \ \ \ \ \ \ \ \ \ \ \ \ \  \ \ \ \ \ \ \ $\Box$

\begin{theorem}\label{2-adic of order d}
Let $p=df+1$ and $q=df^{\prime}+1$ be two odd primes satisfying $\mathrm{gcd}(p-1,q-1)=d$ and $p<q$. Suppose $\{s_i\}_{i=0}^{N-1}$ is the modified Jacobi sequence with period $N=pq$. Then the 2-adic complexity $\phi_2(s)$ of $\{s_i\}_{i=0}^{N-1}$ is bounded by
\begin{equation}
\phi_2(s)\geq pq-p-q-1.
\end{equation}
Specially, if $q=p+2$, then the 2-adic complexity of $\{s_i\}_{i=0}^{N-1}$ is maximal.
\end{theorem}
\noindent{\bf Proof.}
By Lemmas \ref{main method-2} and \ref{main method-3}, we can get
\begin{eqnarray}
\mathrm{det}(A)&=&\prod_{a=0}^{N-1}S(\omega_N^a)\nonumber\\
&=&\prod_{a\in R}S(\omega_N^a)\prod_{a\in D_0^{\ast}}S(\omega_N^a)\prod_{a\in D_1^{\ast}}S(\omega_N^a)\prod_{a\in P}S(\omega_N^a)\prod_{a\in Q}S(\omega_N^a)\nonumber\\
&=&2\left(\frac{p+1}{2}\right)^{q}\left(\frac{q-1}{2}\right)^{p}\left(\Omega_0\Omega_1\right)^{\frac{(p-1)(q-1)}{2}}.\nonumber
\end{eqnarray}
From Theorem \ref{gauss period-thm}, we know that $\Omega_0\Omega_1=\frac{1-pq}{4}$ if $ff^{\prime}$ is odd or if $ff^{\prime}$ is even and $d\equiv0\pmod{4}$, and $\Omega_0\Omega_1=\frac{1+pq}{4}$ if $ff^{\prime}$ is even and $d\equiv2\pmod{4}$. Then we have
\begin{eqnarray}
\mathrm{det}(A)=\left\{
\begin{array}{ll}
2\left(\frac{p+1}{2}\right)^{q}\left(\frac{q-1}{2}\right)^{p}\left(\frac{pq-1}{4}\right)^{\frac{(p-1)(q-1)}{2}},\ \mathrm{if}\ ff^{\prime}\ \mathrm{is\ odd}, \\
\ \ \ \ \ \ \ \ \ \ \ \ \ \ \ \ \ \ \ \ \ \mathrm{or\ if}\ ff^{\prime}\ \mathrm{is\ even\ and}\ d\equiv0\pmod{4}\\
2\left(\frac{p+1}{2}\right)^{q}\left(\frac{q-1}{2}\right)^{p}\left(\frac{pq+1}{4}\right)^{\frac{(p-1)(q-1)}{2}},\ \mathrm{if}\ ff^{\prime}\ \mathrm{is}\\ \ \ \ \ \ \ \ \ \ \ \ \ \ \ \ \ \ \ \ \ \ \mathrm{even\ and}\ d\equiv2\pmod{4}.
\end{array}
\right.
\end{eqnarray}
Now, let $r$ be an any prime factor of $2^{N}-1$ and $\mathrm{Ord}_r(2)$ the multiplicative order of 2 modular $r$. Then we get $\mathrm{Ord}_r(2)|N$. Note that $N=pq$. Therefore, $\mathrm{Ord}_r(2)=pq$, $p$ or $q$. Next, we will prove $\mathrm{gcd}(\mathrm{det}(A),2^{N}-1)\leq(2^p-1)(2^q-1)$. To this end, we need discuss the following three cases.
\begin{itemize}
\item[Case 1.] $\mathrm{Ord}_r(2)=pq$. By Fermat Theorem, we know that $r|2^{r-1}-1$, which implies $\mathrm{Ord}_r(2)\leq r-1$. Since $\mathrm{Ord}_r(2)=pq$, then we have $pq\leq r-1$, i.e., $r\geq pq+1$. But we know that $\frac{p+1}{2}<pq+1$, $\frac{q-1}{2}<pq+1$, and $\frac{pq\pm1}{4}<pq+1$, then $\mathrm{gcd}(\mathrm{det}(A),2^N-1)=1<(2^p-1)(2^q-1)$.
\item[Case 2.] $\mathrm{Ord}_r(2)=p$. Similar argument to that in Case 1, we can get $r\geq p+1$, which implies $\mathrm{gcd}(\frac{p+1}{2}, 2^N-1)=1$. It is obvious that $r|2^p-1$. If $r|\frac{q-1}{2}$ or $r|\frac{pq\mp1}{4}$ (Here $\frac{pq\pm1}{4}$ corresponds to the cases of odd $ff^{\prime}$ and even $ff^{\prime}$), then we have $\mathrm{gcd}(r,q)=1$. Furthermore, by Lemma \ref{num-theory}, we know $\mathrm{gcd}(2^p-1,\frac{2^N-1}{2^p-1})=\mathrm{gcd}(2^p-1,q)$. Thus we have $\mathrm{gcd}(r,\frac{2^N-1}{2^p-1})=1$.
\item[Case 3.] $\mathrm{Ord}_r(2)=q$. It is obvious that $r|2^q-1$. From Lemma \ref{num-theory}, we know that $\mathrm{gcd}(2^q-1,\frac{2^N-1}{2^q-1})=1$, which implies that $\mathrm{gcd}(r,\frac{2^N-1}{2^q-1})=1$.
\end{itemize}
Combining Case 2 and Case 3, no matter $\mathrm{Ord}_r(2)=p$ or $\mathrm{Ord}_r(2)=q$, we always have $\mathrm{gcd}(\mathrm{det}(A),2^N-1)|(2^p-1)(2^q-1)$ and $\mathrm{gcd}(\mathrm{det}(A),\frac{2^N-1}{(2^p-1)(2^q-1)})=1$, which implies that $\mathrm{gcd}(\mathrm{det}(A),2^{N}-1)\leq(2^p-1)(2^q-1)$. From Lemma \ref{main method-1}, we have
$$
\mathrm{gcd}\left(S(2),2^N-1\right)\leq\mathrm{gcd}\left(\mathrm{det}(A),2^N-1\right)\leq(2^p-1)(2^q-1).
$$
Then, by Eq. (\ref{2-adic calculation}), we get $$\phi_2(s)=\left\lfloor\mathrm{log}_2\frac{2^N-1}{\mathrm{gcd}(2^N-1,S(2))}\right\rfloor\geq\left\lfloor\mathrm{log}_2\frac{2^{pq}-1}{(2^p-1)(2^q-1)}\right\rfloor\geq pq-p-q-1.$$

Specially, if $q=p+2$, then we know that $ff^{\prime}$ is even, which implies that $\mathrm{det}(A)=2\left(\frac{p+1}{2}\right)^{p(p+2)+1}$. From the above argument, we know that
$$\mathrm{gcd}\left(\frac{p+1}{2},2^{p(p+2)}-1\right)=1.$$
The desire result follows.
\ \ \ \ \ \ \ \ \ \ \ \ \ \ \ \ \ \ \ \ \ \ \ \ \ \ \ \ \ \ \ \ \ \ \ \ \ \ \ \ \ \ \ \ \ \ \ \ \ \ \ \ \ \ \ \ \ \ \ \ \ \ \ \ \ \ \ \ \ \ \ \ \ \ \ \ \ \ \ \ \ \ \ \ \ \ \ \ \ \ \ \ \ \ \ $\Box$

\section{Summary}
In this paper, we derive gauss periods of a class of generalized cyclotomic sets from Whiteman generalized cyclotomic classes. As an application, a lower bound on the 2-adic complexity of modified Jacobi sequences is determined.
Our result shows that the 2-adic complexity is at least $pq-p-q-1$ with period $N=pq$, which is obviously large enough to resist against RAA for FCSR.
\section*{Acknowledgement}
Parts of this work were written during a very pleasant visit of the first author to the University of Carleton University in
School of Mathematics and Statistics. She wishes to thank the hosts for their hospitality.


\begin{thebibliography}{}

\bibitem{Bai Enjian} Bai, E., Liu, X., Xiao, G.: Linear complexity of new generalized cyclotomic sequences of order two of length $pq$. IEEE Trans. Inform. Theory 51, 1849-1853 (2005).

\bibitem{Cai Han} Cai, H., Liang, H., Tang, X.: Constructions of optimal 2-D optical orthogonal codes via generalized cyclotomic classes. IEEE Trans. Inform. Theory 61, 688-695 (2015).

\bibitem{Davis} Davis, P J.: Circulant Matrices. New York, NY, USA: Chelsea, 1994.

\bibitem{Ding Cunsheng-4} Ding, C., Xing, C.: Several classes of $(2^m-1,w,2)$ optical orthogonal codes. Discrete Applied Mathematics 128, 103-120 (2003).

\bibitem{Ding Cunsheng-5} Ding, C., Xing, C.: Cyclotomic optical orthogonal codes of composite lengths. IEEE Trans. Inform. Theory 52, 263-268 (2004).

\bibitem{Ding Cunsheng} Ding, C.: Cyclotomic constructions of cyclic codes with length being the product of two primes. IEEE Trans. Inform. Theory 58, 2231-2236 (2012).

\bibitem{Ding Cunsheng-3} Ding, C.: Cyclic codes from the two-prime sequences. IEEE Trans. Inform. Theory 58, 3881-3891 (2012).

\bibitem{Ding Cunsheng-1} Ding, C., Helleseth, T.: On the linear complexity of Legendre sequences. IEEE Trans. Inform. Theory 44, 1693-1698 (1998).

\bibitem{Fan Cuiling} Fan, C., Ge, G.: A unified approach to Whiteman's and Ding-Helleseth's generalized cyclotomy over residue classs rings. IEEE Trans. Information Theory 60, 1326-1336 (2014).

\bibitem{Hu Liqin} Hu, L., Yue, Q.: Gauss periods and codebooks from generalized cyclotomic sets of order four. Design, Codes and Cryptography 69, 233-246 (2013).

\bibitem{Li Xiaoping} Li, X., Ma, W., Yan, T., Zhao, X.: Linear complexity of a new generalized cyclotomic sequence of order two of length $pq$. IEICE Transactions 96-A, 1001-1005 (2013).



\bibitem{Massey} Massey,J. L.: Shift-register synthesis and BCH decoding. IEEE Trans. Inform. Theory 15, 122-127 (1969).
\bibitem{Andrew Klapper} Klapper, A., Goresky, M.: Feedback shift registers, 2-adic span, and combiners with memory. Journal of Cryptology  10, 111-147 (1997).


\bibitem{Tang-Ding}Tang, X., Ding, C.: New classes of balanced quaternary and almost balanced binary sequences with optimal autocorrelation Value. IEEE Trans. Inform. Theory 56, 6398-6405 (2010).

\bibitem{Tang-Fan bound}Tang, X., Fan, P., Matsufuji,S.: Lower bounds on the maximum correlation of sequences with low or zero correlation zone. Electron. Lett. 36, 551-552 (2000).

\bibitem{Tian Tian} Tian, T., Qi, W.: 2-Adic complexity of binary $m$-sequences. IEEE Trans. Inform. Theory 56, 450-454 (2010).

\bibitem{Whiteman-1} Whiteman, A L.: A family of difference sets. lllinois Journal of Mathematics 6, 107-121 (1962).

\bibitem{Xiong Hai}Xiong, H., Qu, L., Li, C.: A new method to compute the 2-adic complexity of binary sequences. IEEE Trans. Inform. Theory 60, 2399-2406 (2014).

\bibitem{Xiong Hai-2}Xiong, H., Qu, L., Li, C.: 2-Adic complexity of binary sequences with interleaved structure. Finite Fields and Their Applications 33, 14-28 (2015).
\bibitem{Yan-thesis} Yan, T.: Study on Constructions and Properties of Pseudo-Random Sequence. Ph. D Thesis, 2007.

\bibitem{Yan Tongjiang-1}Yan, T., Du, X., Xiao, G., Huang, X.: Linear complexity of binary Whiteman generalized cyclotomic sequences of order $2^k$. Information Sciences 179, 1019-1023 (2009).

\bibitem{Zeng Xiangyong} Zeng, X.,Cai, H., Tang, X., Yang, Y.: Optimal frequency sequences of odd length. IEEE Trans. Inform. Theory 59, 3237-3248 (2013).

\bibitem{Zeng Xiangyong-2} Xiao, Z.,Zeng, X.: 2-Adic complexity of two classes of generalized cyclotomic binary sequences. International Journal of Foundations of Computer Science 27, 879-893 (2016).

\bibitem{Zhou}Zhou, Z., Tang, X., Gong, G.: A new classes of sequences with zero or low correlation zone based on interleaving technique. IEEE Trans.Inform. Theory 54, 4267-4273 (2008).


\end{thebibliography}
\end{document}